\documentclass[twocolumn,aps,prl,amsfonts, superscriptaddress]{revtex4-1}
\usepackage{amsmath}
\usepackage{graphicx}
\usepackage{bm}
\usepackage{array}
\usepackage{dcolumn}
\usepackage{float}
\usepackage{epstopdf}
\usepackage{lineno}

\newcommand{\ket}[1]{\ensuremath{\left|#1\right\rangle}}
\newcommand{\bra}[1]{\langle{#1}|}
\newcommand{\vc}[1]{\ensuremath{\mathbf{#1}}}
\newcommand{\avg}[2]{\left\langle {#1} \right\rangle_{#2}}
\newcommand{\abs}[1]{|{#1}|}

\usepackage{xcolor}
\usepackage{braket,mleftright}
\begin{document}

\def\d{\downarrow}
\def\u{\uparrow}
\def\e{\mathcal{E}}
\def\ba{\begin{eqnarray}}
\def\ea{\end{eqnarray}}
\def\beq{\begin{equation}}
\def\eeq{\end{equation}}

\title{Which way does stimulated emission go?}
\date{\today}
\author{J. D. Wong-Campos}
\affiliation{Department of Chemistry and Chemical Biology, Harvard University} 
\author{J. V. Porto}
\affiliation{Joint Quantum Institute, NIST and the University of Maryland} 
\author{Adam E. Cohen}
\affiliation{Department of Chemistry and Chemical Biology, Harvard University} 
\affiliation{Department of Physics, Harvard University}
\email[cohen@chemistry.harvard.edu]{}

\date{\today}

\begin{abstract}

Is it possible to form an image using light produced by stimulated emission?  Here we study light scatter off an assembly of excited chromophores.  Due to the Optical Theorem, stimulated emission is necessarily accompanied by excited state Rayleigh scattering.  Both processes can be used to form images, though they have different dependencies on scattering direction, wavelength and chromophore configuration.  Our results suggest several new approaches to optical imaging using fluorophore excited states.
\end{abstract}

\maketitle

In 1916, Einstein introduced the concept of stimulated emission, wherein on-resonance radiation induces an electronically excited atom (or later, molecule) to transition to the resonant lower energy state while simultaneously emitting a quantum of radiation to conserve energy \cite{Einstein1916}. Based on thermodynamic considerations, Einstein deduced that the photon emission must be ``vollst\"andig gerichtete Vorg\"ange" -- fully directed events \cite{Einstein1917}. Subsequent analyses argued that for bulk excited-state media the stimulated emission photons have the same phase, direction, and frequency, \cite{Sargent1974} i.e. they are indistinguishable, from the stimulating photons. Einstein’s description of stimulated emission has been verified by countless laboratory tests, and underlies the working of lasers, atomic clocks and MRI machines.	

Despite the clear successes of this description, a widely held conceptual problem emerges when one considers stimulated emission of an individual excited atom or dye molecule. Consider stimulating light incident on an electronically excited molecule~\cite{Vahala1993}.  According to Maxwell's equations, the molecule radiates in a classical dipole emission pattern. This dipole emission is not ``fully directed" but is symmetric around the transition dipole axis, bearing no relation to the spatial mode of the stimulating light. Using this reasoning, it would be possible to separate dipole radiation from stimulating light with a suitable spatial filter, thus allowing the formation of background-free images from stimulated emission photons alone. In this picture, directional emission is recovered for extended homogeneous media composed of many excited dipoles (e.g. a laser medium) because the emitting dipoles interfere constructively in the forward direction and destructively in all other directions \cite{Murray2018}.  Is stimulated emission of an isolated dipole ``fully directed'', or is it in an isotropic dipole pattern, or is it something else?

\begin{figure*}[t]
    \includegraphics[width=\textwidth]{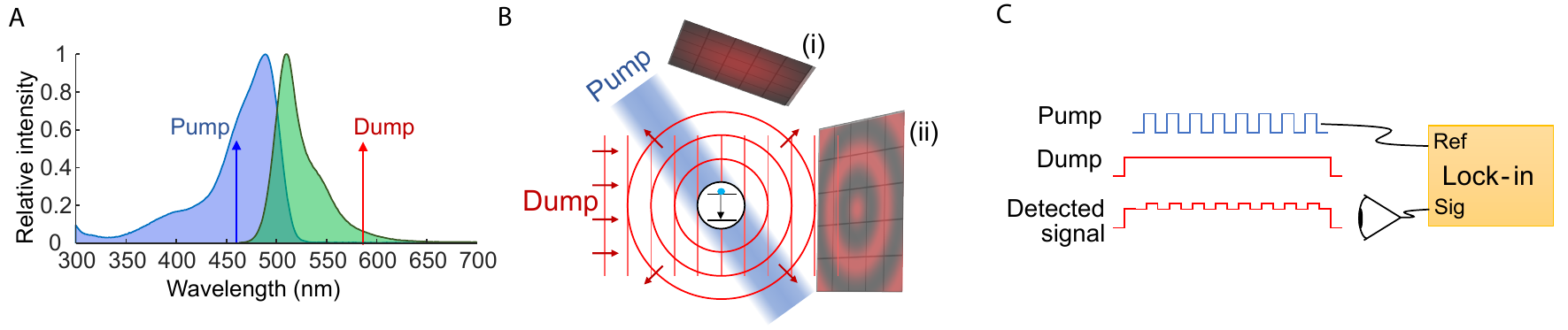}
    \caption{Light scattering off an electronically excited molecule.  A) Excitation and Stokes-shifted emission spectra of eGFP. The pump (blue) excites the molecule. The dump (red) drives stimulated emission and scattering. B)  A Dump plane wave drives stimulated emission of an excited fluorophore.  The stimulated dipolar emission is either detected (i) alone or (ii) via interference with the Dump.  C) Lock-in detection scheme.  An oscillatory pump leads to modulated signal at the dump wavelength.  The lock-in measurement could be done in a point-scanning mode with a conventional lock-in amplifier, or via wide-field illumination and camera-based detection, with lock-in achieved by subtracting camera frames with Pump off from frames with Pump on. }
    \label{fig:exc_em}
\end{figure*}

For clarity in the discussion that follows, we consider fluorescent molecules, which absorb at one wavelength and emit at a longer wavelength (see Fig. \ref{fig:exc_em}). We name the beam that prepares molecules in the excited state the ``pump''.  The local excitation probability due to the pump, multiplied by the local fluorophore density, sets the spatial profile of excited fluorophores. In most of what follows we assume a spatially homogeneous and saturating pump, whose only role is to maintain much of the population in the excited state, i.e. population inverted.  We name the beam that drives stimulated emission the ``dump''. Due to the possibility of interference between the dump and the molecule emission, the coherence properties of the dump are important.

One can imagine an experiment where a spatially heterogeneous assembly of fluorophores, e.g.\ a fluorescently labelled biological sample, is pumped into a population-inverted state, and then exposed to a dump beam tuned to match an emission wavelength (Fig. \ref{fig:exc_em}a).  Are the photons coming off the sample distinguishable from the unscattered dump photons?  Can they be used to form an image?  

Recently, Varma and coworkers performed an elegant experiment along these lines using sub-diffraction fluorescent polystyrene beads or nitrogen-vacancy (N-V) doped nanodiamonds \cite{Varma2020}.  They pumped the sample into an excited state, and then de-excited the sample using a collimated dump beam.  They arranged their optical system in a phase-contrast configuration, wherein the emitted dipole field mixed with a small amount of the phase-shifted unscattered dump to form an image on the camera.  These experiments clearly showed a dipole-like pattern of emission that was coherent with the dump and that could be used to form an image of the sample.  We must now reconcile the unambiguous experimental results of Varma \textit{et al.}\ with the widely held understanding that stimulated emission goes into the spatial mode of the stimulating beam.  The Varma \textit{et al.}\ experiments also raise the question of whether the pure stimulated (i.e.\ coherent) dipole field of the excited fluorophores can be detected in the absence of interference with the dump.

Here we analyze light scatter off sub-wavelength assemblies of excited fluorophores.  Our main findings are: (1) Stimulated emission is always accompanied by excited state Rayleigh scattering.  (2) Stimulated emission and excited state Rayleigh scattering have different angular distributions and different spectral lineshapes.  Both modalities can, in principle be used for image formation, though the Rayleigh scattering signal from single excited molecules is very small. (3) For sub-wavelength assemblies of non-interacting fluorophores, the excited state Rayleigh scattering scales with the number of fluorophores as $N^2$, while stimulated emission scales only as $N$, so for dense samples excited state Rayleigh scattering may become substantial; (4) The subtle question of whether the stimulated emission is directional or isotropic is formally resolved by an appropriate expansion of the dump beam electric field in vector spherical harmonics and calculation of the Poynting vector for dump + scattered electric fields.

This last finding relies on the important distinction between electric fields and energy density. The radiated energy density, which is the observable in an experiment, results from a combination of the excitation field and the scattered field, which are typically in distinct modes. Thus, the concept of a particular ``mode" for the energy density of stimulated emission is not well defined. The spatial distribution of stimulated emission is, however, determined by the overlap of the spatial modes of the incident and scattered fields, so that stimulated emission photons are indistinguishable from the dump photons. Nevertheless, by comparing the profile of the dump with and without first pumping the sample, one can form an image \cite{Min2009}. 

Together, our results reconcile the apparent contradiction between the experimental measurements of Varma \textit{et al.}~\cite{Varma2020} and the widely held understanding that stimulated and stimulating photons are indistinguishable.  The calculations provide a framework for designing imaging systems that use fluorophore excited states to create contrast.  Elements of the discussion below have been presented in previous articles \cite{Cray1982, Vahala1993, takamizawa2012rayleigh, takamizawa2012spatial, Pollnau2019} and books~\cite{Scully1997, Murray2018, Mandel1995}, but to our knowledge these ideas have not been brought together in the context of imaging via light scatter off excited matter.

\subsection{Classical picture}
First we introduce a classical picture of light scattering off excited molecules, where the molecular properties are described by a linear polarizability.  For ground-state molecules, this approach accurately describes absorption and Rayleigh scattering.  For excited-state molecules, the classical picture accurately describes stimulated emission and excited-state Rayleigh scattering, but not spontaneous emission. In a later section we introduce a quantum mechanical description which captures stimulated, Rayleigh, and spontaneous radiation terms.

Consider a continuous-wave electromagnetic field $\vc{E}_i(\vc{r},t) = \vc{E}_i(\vc{r})e^{-i\omega t}$, with an accompanying magnetic field that follows Faraday's law $i\omega\vc{B}_i= \nabla \times\vc{E}_i$.  When the applied field impinges on a dipolar scatterer, the resulting electric field $\vc{E}(\vc{r},t)$ is
\begin{equation}
    \vc{E} = \vc{E}_i+\vc{E}_d
\end{equation}
where $\vc{E}_d(\vc{r},t)$ is the scattered field, and similarly for the magnetic field.  We take the real part of the fields to calculate physical observables and assume that any additional time dependence to $\vc{E}_i(\vc{r})$ is slow enough to satisfy the slowly varying approximation, $|\dot{\vc{E}}_i(\vc{r})|\ll \omega |\vc{E}_i(\vc{r})|$.  

The scattered E- and B-fields arise from the induced oscillating dipole, which at large distances $r$ are given by~\cite{jackson_classical_1999}:
\begin{eqnarray}
    \label{eqn:farE}
    \vc{E}_d & \simeq & \frac{1}{4\pi\epsilon_0}\frac{\omega^2}{c^2 r}\left(\hat{\vc{r}}\times\vc{d}(t^\prime)\times\hat{\vc{r}}\right),\\
    \vc{B}_d & \simeq & \frac{1}{c} \frac{1}{4\pi\epsilon_0}\frac{\omega^2}{c^2 r}\left(\hat{\vc{r}}\times\vc{d}(t^\prime)\right),
\end{eqnarray}
where $\vc{r}=r \hat{\vc{r}}$, $\vc{d}(t^\prime)=\vc{d}_s e^{-i\omega t^\prime}$ is evaluated at the retarded time $t^\prime = (t-r/c)$ and we choose the origin to be at the position of the dipole. We also assume that $\vc{d}_s$ satisfies the slowly varying approximation $|\dot{\vc{d}}_s|\ll \omega |\vc{d}_s|$.

The time averaged energy flow is determined by the Poynting vector $\vc{S} =(1/2 \mu_0)\Re[ \vc{E} \times  \vc{B}^*$]. The difference between the Poynting vector of the total field  $\vc{S}_\textrm{tot}$ and the Poynting vector of the incident field, $\vc{S}_i,$ is~\cite{Vahala1993,Cray1982,Berg2008,Striebel2017}    
\begin{eqnarray}
\label{eqn:excessS}
    \Delta \vc{S} &=& \vc{S}_{\mathrm{tot}}-\vc{S}_i \nonumber \\
    &=& \vc{S}_{\mathrm{cross}}+\vc{S}_d,
\end{eqnarray}
where
\begin{eqnarray}
\vc{S}_{\mathrm{cross}}&=& \frac{1}{2\mu_0}\Re[\vc{E}_i\times \vc{B}^*_d + \vc{E}_d \times \vc{B}^*_i], \label{eq:Scross} \\ 
\vc{S}_{d}&=&
    \frac{1}{2\mu_0}\Re[\vc{E}_d \times \vc{B}^*_d] \label{eq:Sdip}
\end{eqnarray}
and $\Re$ indicates the real part.  The integral of $\Delta \vc{S}
\cdot \hat{\vc{r}}$ over a far-field surface of radius $r$ gives the power added to or removed from the total field, $P_\text{abs} = \int r^2 \Delta \vc{S}\cdot \hat{\vc{r}}\  d\Omega$ following Poynting's theorem (see Appendix). By energy conservation, $P_\text{abs}$ equals the negative of the rate of change of the internal energy of the molecule, $-P_\text{mol}$, and for purely elastic scattering $P_\text{abs}=0$.

The cross term, Eq.~\ref{eq:Scross},  represents interference of the scattered field and the incident field.  Clearly this term can only exist in regions where the unscattered incident field is non-zero, so it cannot be used for background-free detection.  In a transmitted light experiment, the interference term represents the extinction (or in the case of stimulated emission, the amplification), of the incident beam~\cite{Cray1982}.  Specifically, the power added to or removed from the incident field (the `extinction') is given by $P_\text{ext} = \int r^2\vc{S}_\mathrm{cross}\cdot\hat{\vc{r}}\ d\Omega$, where the integral is over a far-field surface.

The second term, Eq.~\ref{eq:Sdip}, represents Rayleigh scattering. The integral of $\vc{S}_d$ gives the scattering power, $P_{\text{R}}= \int r^2\vc{S}_d\cdot\hat{\vc{r}}\ d\Omega$.  In a Rayleigh scattering process the internal energy of the molecule does not change.  Rather, energy is redirected from the incident beam into a dipole radiation pattern.  

In the classical theory where spontaneous emission is not described, neither term in Eq.\ \ref{eqn:excessS} can exist without the other, and extinction (or amplification) of the incoming beam is necessarily accompanied by Rayleigh radiation in a dipole pattern. Equation~\ref{eqn:excessS} implies $P_\text{abs}=P_\text{ext}+P_\text{R}$, a result related to the optical theorem~\cite{newton1976optical}. The quantities $P_\text{abs}$ and $P_\text{ext}$ can be either positive or negative, but $P_\text{R}$ is always positive.  

\subsection{Molecular polarizabilities}
For an isolated molecule, the induced dipole moment $\vc{d}$ is related to the incident electric field at the dipole by $\vc{d} = \bm{\alpha}\vc{E}_i$, where $\bm{\alpha} = \bm{\alpha}' + i \bm{\alpha}''$ is the complex polarizability tensor.  We will assume an isotropic distribution of molecules, which makes $\bm{\alpha}$ a scalar, $\alpha = \frac{1}{3}\textrm{Tr}(\bm{\alpha})$. The quantities $\alpha'$ and $\alpha''$ can be calculated at varying levels of theory \cite{mukamel1999principles} or determined experimentally.  Here we describe how to calculate $\alpha'$ and $\alpha''$ for ground and excited states in terms of experimentally measurable quantities.

The ground state spectrum of $\alpha''$ can be calculated from an optical extinction spectrum, as follows.  The power extracted from the incident field is $P_{\mathrm{d}} = \avg{\vc{E}_i(0)\cdot \dot{\vc{d}}}{t}$.  Here $P_{\mathrm{d}}$ represents the sum of the molecular excitation and Rayleigh scattering.  In this product, only the contribution due to $\alpha''$ has non-zero time-average.  The quantity $\alpha''$ is typically associated with molecular excitation, but it also contributes to scattering via a contribution from radiative damping.  Thus $P_\mathrm{d} = -P_\text{ext}=P_\text{mol}+P_\text{R}$, where $P_\text{mol}=-P_\text{abs}$ is the power absorbed by the molecule. 

The average intensity (power per area) of the incident beam is $I_i = \frac{1}{2} \epsilon_0 c \abs{\vc{E}_i}^2$.  The total extinction cross section of the molecule, $\sigma_\text{tot} = P_\mathrm{d}/I_i$, is therefore:
\begin{equation}
\label{eqn:sigmaAbs}
    \sigma_\text{tot} = \frac{2\pi}{\lambda\epsilon_0 }\alpha''.
\end{equation}
The experimentally measured decadic molar extinction coefficient  $\epsilon_{\textrm{ext}}(\lambda)$ (units $\textrm{molar}^{-1} \textrm{cm}^{-1}$) and extinction cross section (units $\textrm{m}^2$) are directly proportional, related by:
\begin{equation}
\label{eqn:eps2sigma}
    \sigma_{\text{tot}}(\lambda) = \frac{2.303 \epsilon_{\mathrm{ext}}(\lambda)}{10 N_A},
\end{equation}
where $N_A$ is Avogadro's number.  Eqs.~\ref{eqn:eps2sigma} and \ref{eqn:sigmaAbs} together relate $\alpha''$ to $\epsilon_{\mathrm{ext}}$ which can be measured in a spectrometer.

To calculate $\alpha''$ in the excited state, we define the normalized emission spectrum $g(f)$, where $f$ is the emission frequency and $\int g(f)df = 1$.  The stimulated emission cross section is expected to be proportional to $g(f)$.  One can obtain the proportionality factor from the radiative lifetime, an experimentally measurable quantity.  Specifically, the magnitude of the stimulated emission cross-section is given by~\cite{hilborn2002einstein}:
\begin{equation}
\label{eqn:A2sigma}
    \sigma_{\text{stim}}(f) = \frac{\gamma_r\lambda^2}{8 \pi}g(f),
\end{equation}
where $\gamma_r$ radiative decay rate constant.  In fluorescent molecules, radiative decay often competes with nonradiative pathways, so $\gamma_r$  can be determined experimentally from the fluorescence quantum yield (QY) and the total lifetime, $\tau$ of the excited state via $A = \textrm{QY}/\tau$.  Alternatively, $\gamma_r$ can be estimated from the absorption spectrum via the Strickler-Berg equation~\cite{strickler1962relationship}.

To convert from the stimulated emission cross section, Eq.~\ref{eqn:A2sigma}, to the excited-state polarizability, $\alpha''$, a negative sign is introduced into Eq.~\ref{eqn:sigmaAbs} to account for the increase of energy in the incident beam in stimulated emission~\cite{lax1952franck}.  Here we assume there are no resonant higher-lying excited states to which the molecule can be further excited.

In either the ground or excited states, the real part of the polarizability, $\alpha'$, can be calculated from $\alpha''$ via the Kramers-Kronig relations:
\begin{equation}
    \label{eqn:KK}
    \alpha'(\lambda) = \frac{2 \lambda^2}{\pi}\mathcal{P}\int_0^{\infty}\frac{\alpha''(\lambda')}{\lambda'(\lambda^2 - \lambda'^2)}d\lambda',
\end{equation}
where $\mathcal{P}$ indicates the Cauchy principal part of the integral.  Given the sign change in $\alpha''$ between ground and excited states, Eq.~\ref{eqn:KK} implies that there is also a sign change in $\alpha'$.

\subsection{Measured lineshapes}

\begin{figure*}[t]
    \centering
    \includegraphics[width=2.1\columnwidth]{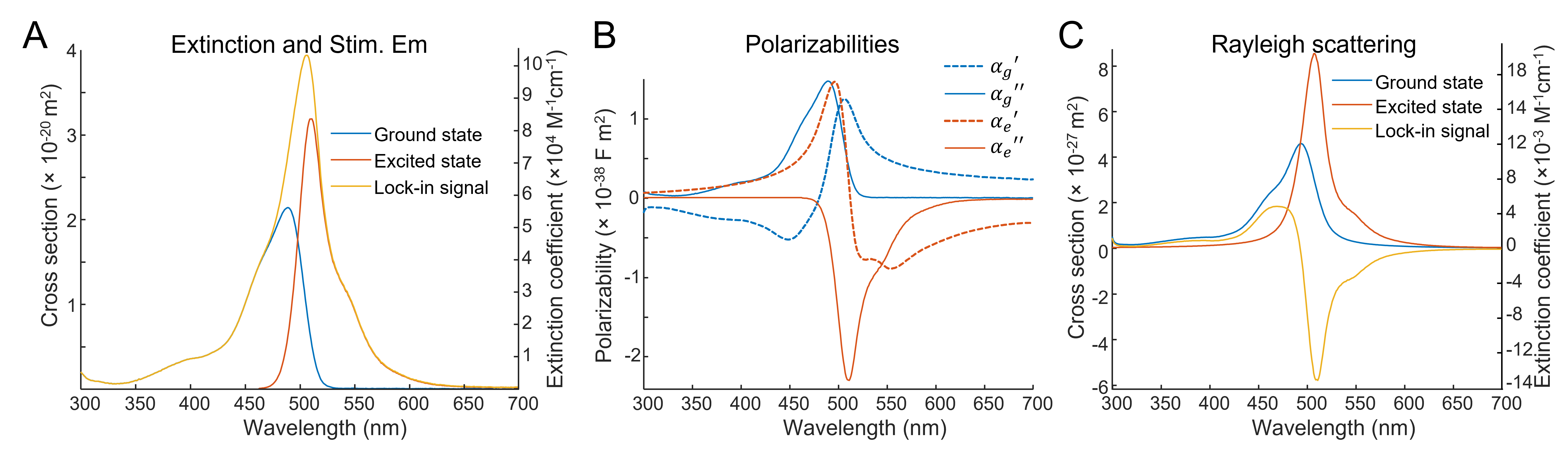}  
    \caption{Lineshapes for excited-state imaging of GFP.  A) Extinction and stimulated emission spectra scaled to show respective cross sections.  A lock-in measurement of the transmitted dump beam as a function of dump wavelength, at fixed pump wavelength, is proportional to the sum of the extinction and emission spectra.  B) Polarizabilities of the ground and excited states calculated using eqs.~\ref{eqn:sigmaAbs}, \ref{eqn:eps2sigma} and \ref{eqn:KK}. C) Rayleigh scattering spectra.  A lock-in measurement is proportional to the difference between the ground and excited state scattering spectra.  In (A) and (C) the magnitudes of the lock-in signals will be proportional to the excitation probability per molecule and should not be referenced to the vertical axes.}
    \label{fig:GFP}
\end{figure*}

There are two ways to configure an imaging experiment with these concepts.  One approach is to mix a small amount of the phase-shifted dump with the scattered field from the molecule. In this case the contrast is dominated by an interference term as in Eq.~\ref{eq:Scross}, which is connected to the material properties by Eq.~\ref{eqn:sigmaAbs}. This was the approach followed by Varma and coworkers~\cite{Varma2020}.

Alternatively, one could arrange a detector with appropriate spatial and spectral filters to avoid collecting any of the unscattered dump and spontaneous emission.  In this case the contrast would come purely from Rayleigh scattering, Eq.~\ref{eq:Sdip}. The far-field Rayleigh scattered intensity distribution is $I_R = \vc{S}_d\cdot \hat{\vc{r}} = \frac{1}{2}\epsilon_0 c \abs{\vc{E}_d}^2$ (see Appendix), and the Rayleigh scattering cross section is $\sigma_R = P_R/I_i$, which evaluates to:
\begin{equation}
    \label{eqn:sigmaR}
    \sigma_R = \frac{8 \pi^3}{3 \epsilon_0^2 \lambda^4}\abs{\alpha}^2.
\end{equation}
For molecules with a Lorentzian absorption peak, the extinction spectrum (Eq.~\ref{eqn:sigmaAbs}) and the Rayleigh scattering spectrum (Eq.~\ref{eqn:sigmaR}) have the same lineshape, but in general for non-Lorentzian absorption peaks, extinction and Rayleigh spectra can be different from each other.  Furthermore, due to the Stokes shift between ground-state absorption and excited-state emission, each type of spectrum will also differ between ground and excited state molecules.

To detect small signals from the excited state population, one would likely use a lock-in detection approach, where the pump at a fixed wavelength is modulated and one looks for synchronous modulation at the dump wavelength (Fig. \ref{fig:exc_em}c).  Thus the image reflects the difference in signal between the excited and ground state fluorophores.

Fig.~\ref{fig:GFP} shows an example of the predicted spectra for enhanced green fluorescent protein (eGFP).  Starting with absorption and emission spectra, we calculated the lineshapes for lock-in measurements of transmitted and Rayleigh scattered light. For a modulated pump at a fixed wavelength of the absorption spectra,  molecules in the ground state attenuate the dump, while molecules in the excited state increase the dump signal at each modulation cycle.  Consequently, the lineshape of the lock-in signal for transmitted light is the sum of the absorption and emission spectra.  For Rayleigh scattering, both the ground state and the excited state contribute positively to the signal, so the lineshape of the lock-in Rayleigh scattering spectrum is proportional to the difference between ground and excited state Rayleigh spectra.

\subsection{Density dependence}

Stimulated emission and excited state Rayleigh scattering have different dependencies on the density of fluorophores.  This is most apparent by examining Eqs.~\ref{eq:Scross} and \ref{eq:Sdip} or Eqs.~\ref{eqn:sigmaAbs} and \ref{eqn:sigmaR}.  The stimulated emission signal is proportional to $\vc{E}_d$, while the Rayleigh signal is proportional to $\abs{\vc{E}_d}^2$ in the far field.  For $N$ fluorophores whose spacing is large compared to the F\"orster radius but small compared to the wavelength, dipole-dipole interactions can be disregarded and the fields add coherently, leading to a $\vc{E}_d$ proportional to $N$.  One thus expects the stimulated emission signal to scale proportional to $N$ and the Rayleigh signal to scale proportional to $N^2$.  For dense samples and sufficiently large detuning, this difference in density dependence can partially compensate for the $\approx10^6$-fold difference in cross sections between stimulated emission and excited state Rayleigh scattering (compare y-axis scales on Figs.~\ref{fig:GFP}A and \ref{fig:GFP}C).  

For molecules freely diffusing in solution, the same arguments that give the concentration dependence of ground-state Rayleigh scattering also apply to excited-state scattering~\cite{loudon2000quantum}.  Specifically, since the fluctuations in density are related to concentration $C$ as $C^{1/2}$, the Rayleigh signal from a homogeneously pumped sample is proportional to $C$. Thus the free-solution Rayleigh signal has the same concentration dependence as the absorption signal, for both ground and excited state Rayleigh scattering.  

\subsection{Intensity dependence}
One might intuitively think that by increasing dump power, one could arbitrarily increase the power into the excited state Rayleigh scattering.  Here we show by classical argument that this is not the case.  For a molecule initially in the excited state, the power into Rayleigh scattering is $P_R = I_i \sigma_R$.  However, the amount of time during which this power is radiated scales inversely with $I_i$ due to driven de-excitation (assuming here that stimulated emission is much faster than spontaneous decay).  The stimulated emission-dominated excited-state lifetime is $\tau_{\mathrm{stim}} = \hbar \omega/\sigma_{\mathrm{stim}} I_i$. Thus the mean energy into Rayleigh scattering per excitation-stimulated emission cycle is bounded by $U_R = \hbar \omega \sigma_R/\sigma_{\mathrm{stim}}$, which is independent of $I_i$. Valhala used a similar scaling argument to establish a bound on the stimulated emission power for an isolated atom undergoing coherent Rabi flopping~\cite{Vahala1993}.

The ratio of powers into Rayleigh scattering vs.\ stimulated emission depends on the dump wavelength because $\sigma_R$ and $\sigma_{\mathrm{stim}}$ have different spectra (see Fig.\ \ref{fig:GFP}).  Thus one may enhance this ratio by suitable choice of dump wavelength. Spontaneous emission further lowers $U_R$ by providing an alternate decay pathway.


\subsection{Where does stimulated emission go?}

\begin{figure*}[t]
    \centering
    \includegraphics{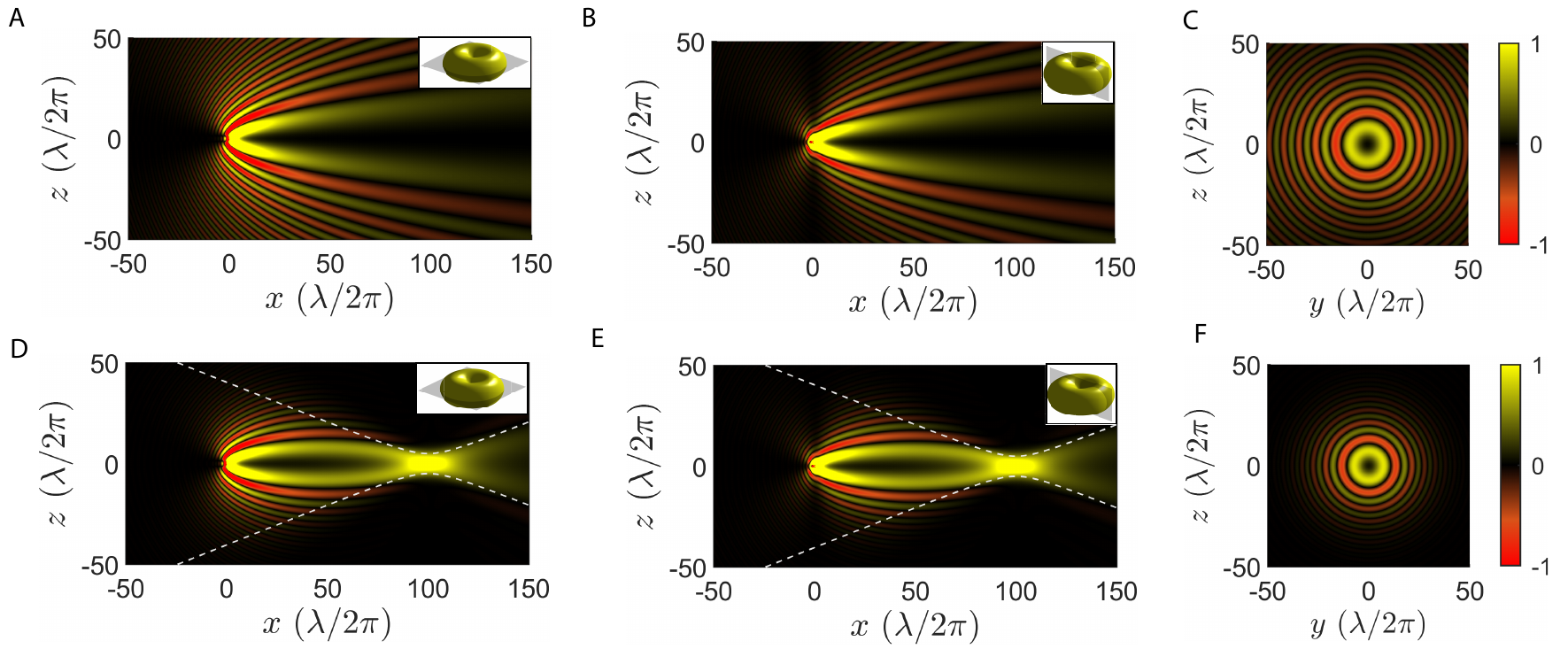}
    \caption{Time-averaged component of the Poynting vector component along the propagation axis associated with the interference of the incident and scattered fields, $\vc{S}_\textrm{cross}$ (Eq.~\ref{eq:Scross}).  A-C) Incident plane wave polarized along the dipole oscillation (z-axis) interferes with scattered dipolar radiation (assuming $\alpha' = 0$, $\alpha'' < 0$). The excess energy density is forward directed, but with fringe structure and zero stimulated emission directly on axis. Insets depict the observation planes (A) perpendicular and (B) parallel to the dipole axis.  (C) Transverse distribution for $x=25$ (in $\lambda/2\pi$ units). D-F) Same as (A-C) with  a Gaussian beam focused at $x=100 \lambda/2\pi$ past the dipole. Dotted lines depict the Gaussian waist of the incident beam.  The interference energy is largely contained within the boundaries of the incident beam. Note that while the plane wave dump evokes zero stimulated emission on axis, the Gaussian dump evokes positive stimulated emission on axis at the Gaussian focus due to the Gouy phase.}
    \label{fig:spatial_stim}
\end{figure*}

It is sometimes said that stimulated emission goes into the ``same mode" as the light that induces the emission. While an electric or magnetic field can be described as occupying modes in a particular basis, the spatial distribution of stimulated emission represents the excess {\em energy} associated with dipole emission. This excess in energy is determined by the contribution of the incident and dipolar fields to the interference terms $\vc{E}_i\times \vc{B}^*_d$ and $ \vc{E}_d \times \vc{B}^*_i$ of eq. (\ref{eq:Scross}), and there is not a well defined ``mode" for stimulated emission. Certainly, the fields must have some mode overlap to interfere, and $\vc{S}_{\text{cross}}$ vanishes wherever either $\vc{E}_i$ or $\vc{E}_d$ vanishes. The detailed spatial structure, however, depends on the amplitudes and relative phase fronts of both fields. The dipole field is given by Eq.~\ref{eqn:farE}, but $\vc{E}_i$ can be any solution to Maxwell's equations.  

Examples of the cross interference patterns for two different $\vc{E}_i$ are shown in Fig.~\ref{fig:spatial_stim}. For a plane wave dump beam, the time averaged excess stimulated energy density is largely in the plane wave direction, but has interference fringe structure and is not equivalent to the (constant) energy density of the stimulating plane wave. Indeed, the optical theorem for purely imaginary $\alpha$ (i.e.\ when $\lambda_{\text{dump}}$ is exactly on resonance) implies that the directly forward scattered light is 90$^{\circ}$ out of phase with $\vc{E}_i$ so that the time averaged interference term vanishes directly in the forward direction (Fig.~\ref{fig:spatial_stim}).

Next we consider a converging Gaussian beam whose focus is displaced somewhat beyond an excited fluorophore (Fig.~\ref{fig:spatial_stim} d-f).  Again we consider a purely imaginary $\alpha$, i.e.\ on-resonance excitation.  In this case the excess stimulated energy density has a local maximum at the focus of $\vc{E}_i$. The 90$^{\circ}$ Gouy phase shift at the Gaussian focus of the incident beam compensates for the 90$^{\circ}$ phase shift between $\vc{E}_i$ and $\vc{E}_d$, leading to constructive interference at the Gaussian focus.

\subsection{Where does excited state Rayleigh scattering go?}
In both of the above examples, the excited-state Rayleigh scattering was a pure dipolar radiation pattern.  However, in extended media, the excited-state Rayleigh scattering can have more complicated spatial structures due to its coherent nature.  For example, excited state Rayleigh scattering could be made directional via a transient grating arrangement.  Consider two pump beams of equal frequency,  intensity, and polarization, with wavevectors $\vc{k}_1$ and $\vc{k}_2$.  Their interference creates an intensity grating defined by wavevector $\vc{q}=\vc{k}_1 - \vc{k}_2$.  The wavelength of the intensity grating is \cite{Deeg1995}:
\begin{eqnarray}
  \Lambda & = & 2\pi/q \nonumber \\
  & = & \lambda_{\text{pump}}/(2 \sin \theta_m)
\end{eqnarray}
where $\theta_m$ is the half-angle between the pump beams in the medium, and $\lambda_{\text{pump}} = \lambda_{\text{pump}}^0/n$ is the pump wavelength in the medium, $\lambda_{\text{pump}}^0$ is the free-space wavelength and $n$ is the refractive index of the medium.

Diffraction of the dump beam occurs when it satisfies the Bragg condition,
\begin{equation}
    \sin \theta = \frac{m\lambda_{\text{pump}}}{2\Lambda},
\end{equation}
where $\theta$ is the angle between the dump beam and the normal to the plane of the transient grating, $\lambda_{\text{pump}}$ is the dump wavelength in the medium, and $m$ is an integer.  Here we assume that the grating thickness, $d$, is much greater than $\Lambda$.

If the solution is sufficiently dilute, then the change in complex dielectric constant due to optical excitation of the fluorophores is small.  The Clausius-Mossotti relation then implies that the change in complex dielectric constant, $\Delta \epsilon$ is:
\begin{equation}
    \frac{\Delta \epsilon}{\epsilon_w + 2} = \frac{N_\text{exc}\Delta \alpha}{3 \epsilon_0},
\end{equation}
where $\epsilon_w$ is the optical-frequency dielectric constant of the solvent (here assumed to be water), $N_\text{exc}$ is the number density of excited-state fluorophores, and $\Delta \alpha$ is the difference in molecular polarizability between excited and ground states.  If one further assumes that the pump beams are far from saturating the transition to the excited states, then the induced grating has a sinusoidal spatial dependence, and the diffraction efficiency can be shown to be \cite{Deeg1995}:
\begin{eqnarray}
    \eta & = & \frac{I_\text{diff}}{I_\text{in}} \nonumber \\
    & = & e^{\frac{-\beta_0 d}{\cos\theta}}\left[\sin^2\frac{k\Delta\epsilon'd}{4\epsilon'_w\cos\theta}+\sinh^2\frac{k\Delta\epsilon''d}{4\epsilon'_w \cos\theta} \right],
\end{eqnarray}
where $\beta_0$ is the extinction coefficient of the non-excited medium and $k=2\pi/\lambda_{dump}$.
Assuming that the arguments of the $\sin$ and $\sinh$ functions are small, and neglecting numerical constants, this result simplifies to:
\begin{equation}
\label{eq:grating}
    \eta \propto C^2\left[(\Delta \alpha')^2 + (\Delta \alpha'')^2\right].
\end{equation}
 Here the refractive index contrast of the grating is proportional to dye concentration, $C$, so the intensity of the Bragg-scattered dump beam is proportional to $C^2$.  This approach has been applied to a variety of dye solutions \cite{eichler1977laser}.
 
 Eq.\ (\ref{eq:grating}) has the same spectral dependence as the lock-in detection scheme plotted in Fig.\ \ref{fig:GFP}C.  Indeed, the transient grating experiment can be thought of a spatial analogue of a time-domain lock-in experiment, wherein the diffraction arises from the contrast between the polarizability of the ground-state molecules and the excited-state molecules.

\subsection*{Quantum picture}
The quantum description in the semiclassical approximation involves replacing $\vc{d}$ and $\vc{E}_d$ with operators $\hat{\vc{d}}$ and $\hat{\vc{E}}_d$, while treating $\vc{E}_i$ classically. After solving for the quantum dynamics of $\hat{\vc{d}}$, the radiated field from a transition between states $|j\rangle \rightarrow |i\rangle$ is given by the quantum version of Eq.~\ref{eqn:farE}~\cite{Scully1997}
\begin{equation}
\label{eqn:quantdip}
    \hat{\vc{E}}_d(\vc{r},t) = \frac{1}{2 \pi \epsilon_0} \frac{\omega^2}{c^2 r} \left(\hat{\vc{r}} \times  \hat{\vc{d}}_-(t^\prime) \times \hat{\vc{r}}\right),
\end{equation}
where $t^\prime = t-r/c$, $\hat{\vc{E}}_d(\vc{r},t)$ is the outgoing radiated field, and $\hat{\vc{d}}_-(t)=\vc{d}_{ij} \hat{\sigma}_-(t)$, with $\hat{\sigma}_-=|i\rangle \langle j|$. Here, $\vc{d}_{ij} = \bra{j} e \vc{r} \ket{i}$ is the dipole matrix element. In a complex multilevel system, the fields from multiple transitions are summed. In the appendix we show that the far-field integral in $P_\text{abs} = \int r^2 \Delta \vc{S}\cdot \hat{\vc{r}}\  d\Omega$ is equal to the time average integral of the excess energy density, $ \int r^2 c \Delta u \  d\Omega$, where $\Delta u =  \epsilon_0( \vc{E}_i \cdot \vc{E}^*_d+ \frac{1}{2}|\vc{E}_d|^2) $. Focusing therefore on the electric field, the quantum excess energy density is given by the classical expression with $\vc{E}_d$ replaced with expectation values of $\hat{\vc{E}}_d$,
\begin{equation}
\label{eqn:quantexcessE}
    \langle \Delta u \rangle = \epsilon_0 \vc{E}_i\cdot\langle\hat{\vc{E}}_d\rangle + \frac{\epsilon_0}{2}\langle {\hat{\vc{E}}_d}^\dagger \hat{\vc{E}}_d\rangle. 
\end{equation}

The key differences between the quantum and classical description are 1) quantum mechanics places limits on the magnitude of $\hat{\vc{E}}_d$ since the largest possible dipole moment $\langle \hat{\vc{d}}(t) \rangle $ is bounded by $\vc{d}_{ij}$, 2) the quantum field gives rise to an incoherent fluorescence contribution to the dipole radiation that is independent of $\vc{E}_i$ and 3) since $\langle \hat{\vc{E}}_d^\dagger \rangle \langle \hat{\vc{E}}_d \rangle \leq \langle \hat{\vc{E}}_d^\dagger \hat{\vc{E}}_d\rangle$, the coherent,  classical Rayleigh scattering,  $P_R \propto | \langle \hat{\vc{E}}_d \rangle|^2$, differs from the total dipole radiation (Rayleigh plus fluorescence), $P_{4 \pi} \propto \langle | \hat{\vc{E}}_d|^2\rangle$, and makes up a fraction of the dipole radiation, $P_R/P_{4 \pi}\leq 1$. For fluorophore molecules, this fraction is typically much smaller than one.

The maximum value of $\langle | \hat{\vc{E}}_d|^2\rangle$ is obtained for an undriven, fully inverted state, and in general is proportional to the excited state population. The molecule cannot emit photons into the dipole pattern faster than this fluorescence rate, which is given by the inverse radiative lifetime $\gamma_r=\tau_{r}^{-1}$ with $\tau_r$ as the spontaneous radiative lifetime. The increased radiation by a molecule driven to de-excite during stimulated emission results entirely from the interference terms, since the total energy in the cross terms can be increased at fixed $\vc{E}_d$ by increasing $\vc{E}_i$. Driving an inverted molecule harder will not increase the total dipole-radiated power. This is in contrast to the classical description, or a non-inverted unsaturated dipole, where $\vc{E}_d \propto \vc{E}_i$ and driving harder increases $P_R$. If one merely collects photons from the dipole radiation field while avoiding the dump beam, the simple fluorescence rate is the upper limit on the detection rate. 

The coherent {\em fraction} of dipole radiation, on the other hand, does depend on $\vc{E}_i$. Although we argued above that the inverted state Rayleigh scattering is weak, it is coherent with the dump beam, and is therefore amenable to spectral filtering and phase contrast imagine techniques by interfering the coherently scattered light with phase shifted dump light.

\section{Conclusion}

We have discussed two distinct ways to form images with excited fluorophores. Images formed by stimulated emission are physically similar to simple absorption measurements, but with a reversed sign on the direction of energy transfer between the beam and the sample.  Stimulated emission images are best formed with lock-in methods due to the high background from the dump beam. 

The dipolar field of excited molecules also produces non-directional emission.  In the absence of a dump beam, the dipolar emission is regular fluorescence.  A dump beam can convert some of that dipolar emission into coherent excited-state Rayleigh scattering.  From a practical perspective, the excited state Rayleigh scattering has the merits of coherence and narrow spectrum, allowing tight spectral filtering to remove broad-spectrum background light.  On the other hand, the excited state Rayleigh scattering has the same spectrum as the dump beam, so background from dump scatter off refractive heterogeneities in the sample may be overwhelming and difficult to remove.  As in the case of stimulated emission, lock-in measurements may distinguish between excited state vs. ground state scattering.

An ensemble of chromophores localized within a sub-wavelength volume can enhance coherent isotropic emission by constructive interference. Detection of this coherent component would be similar to dark-field microscopy of non-fluorescent objects such as gold nanoparticles, where the scattering is theoretically unbounded and allows interferometric measurements \cite{Taylor2019}. 

Finally, the change in Rayleigh scattering between ground and excited states can also be understood as a change in refractive index, which for sufficiently diluted samples of fluorescent molecules in a dielectric media, takes the form of $\Delta n = N\Delta \alpha /2\epsilon_0$, where $N$ is the number of molecules and $\Delta \alpha$ is the change in polarizability related to the ground and excited states.


\section{Acknowledgements}
We thank Andrew York for helpful discussions.  AEC is supported by a Vannevar Bush Faculty Fellowship. DWC is a Merck Awardee of the Life Sciences Research Foundation.

\appendix
\section{APPENDIX}

\subsection{Vector spherical functions and dipole scattering    }

Here we show that the total power added to or removed from the incident beam (i.e.\ the integral of Eq.~\ref{eqn:excessS} over a spherical surface) 
depends only on $\vc{E}_i(0)$ for any $\vc{E}_i(\vc{r})$.  Further, an expansion of $\vc{E}_i(\vc{r})$ in the vector spherical function (VSF) basis shows that a single term in that expansion is completely responsible for the integral of $\vc{S}_{\mathrm{cross}}$ (Eq.~\ref{eq:Scross}) over a spherical surface, and the interference of the dipole field with each other term in the expansion integrates to zero.

The VSF's are three-component spherical solutions to Maxwell's equations in Helmholtz form, labeled by two angular momentum numbers $l$ and $m$, and wavevector $k=2\pi/\lambda$: $\vc{N}_{klm}(\vc{r})$ and $\vc{M}_{klm}(\vc{r})$. A pair of transverse $\vc{E}$ and $\vc{B}$ fields centred on $\omega = c k$ can be written as the real part of a sum over VSFs
\begin{eqnarray}
    \label{eqn:Eexpansion}
    \vc{E}(\vc{r},t) &=& \sum_{lm}\left(^{1\! \!}E_{lm} \vc{N}_{klm}( \vc{r})+^{2\! \! \!}E_{lm} \vc{M}_{klm}( \vc{r})\right) e^{- i \omega t}\\
\vc{B}(\vc{r},t)&=& \frac{-i}{c}\sum_{lm}  \left(^{1\! \!}E_{lm} \vc{M}_{klm}( \vc{r})+^{2\! \! \!}E_{lm} \vc{N}_{klm}( \vc{r})\right) e^{- i \omega t}, \label{eqn:Bexpansion}
\end{eqnarray}
where $^{1\! \!}E_{lm}(t^\prime)$ and $^{2\! \!}E_{lm}(t^\prime)$ are expansion coefficients, which are assumed to be slowly varying functions of $t^\prime$. There are source-free VSF solutions, $\vc{N}_{klm}(\vc{r})$ and $\vc{M}_{klm}(\vc{r})$, relevant for describing $\vc{E}_i$ and sourced solutions, $^{\pm}\vc{N}_{klm}(\vc{r})$ and $^{\pm}\vc{M}_{klm}(\vc{r})$, appropriate for a field with a point source ($+$) or sink ($-$), such as $\vc{E}_d$. The source-free and source/sink solutions are related by $\vc{N}_{klm}(\vc{r})=(^{+}\vc{N}_{klm}(\vc{r})+^{-\! \!}\vc{N}_{klm}(\vc{r}))/2.$

The VSFs satisfy a number of orthogonality relations:
local orthogonality and curl relations
\begin{eqnarray}
    \vc{N}_{k l m}(\vc{r})\cdot \vc{M}_{klm}(\vc{r})&=&0,\\
    \nabla \times \vc{N}_{k l m}(\vc{r}) &=& k \vc{M}_{klm}(\vc{r}), \\
    \nabla \times \vc{M}_{k l m}(\vc{r}) &=& k \vc{N}_{klm}(\vc{r}),
\end{eqnarray}
and  orthogonality of dot- and cross-products integrated over solid angle in the far field, $r \rightarrow \infty$,
\begin{eqnarray}
\label{eqn:dotprodorth}
\int \vc{N}_{kl^\prime m^\prime}(\vc{r})\cdot \vc{N}^*_{klm}(\vc{r}) d\Omega & = & A_N \delta_{l^\prime l} \delta_{m^\prime m} \\
\int \vc{M}_{kl^\prime m^\prime}(\vc{r})\cdot \vc{M}^*_{klm}(\vc{r}) d\Omega & = & A_M \delta_{l^\prime l} \delta_{m^\prime m} \\
\int \vc{N}_{kl^\prime m^\prime}(\vc{r})\times (i  \vc{M}^*_{klm}(\vc{r})) \cdot \hat{\vc{r}}\ d\Omega & = &C \delta_{l^\prime l} \delta_{m^\prime m}
\label{eqn:crossprodorth}
\end{eqnarray}
where it is understood that we take the real part for the physical solution. The normalisation depends on whether the functions are source-free and/or sourced. For the source-free case $\vc{N}_{klm}$ (or $\vc{M}_{klm}$):
\begin{eqnarray}
\label{eq:NormNN}
    A_N  (A_M) &=& \begin{cases} 
    \frac{\sin^2(k r)}{(k r)^2}, & l\ \text{odd (even)} \\
    \frac{\cos^2(k r)}{(k r)^2}, & l\ \text{even (odd)}
    \end{cases}\nonumber \\
    C &=& 0.
\end{eqnarray}
For the case where both terms are sourced (e.g. 
\hbox{$^{+\! \!}\vc{N}_{klm}\cdot^{+\! \!}\vc{N}_{k l^\prime m^\prime}$} ):
\begin{equation}
\label{eq:sourcednorm}
    A_N =A_M =C =
    \frac{1}{(k r)^2}.
\end{equation}
For the mixed sourced and source-free case appropriate for the cross terms, (e.g. $\vc{N}_{klm}\cdot ^{+\! \!}\vc{N}_{k l^\prime m^\prime}$ )
\begin{eqnarray}
\label{eq:mixednorm}
    A_N\ \mathrm{or}\ C \ \ (A_M) &=& \begin{cases} 
    \frac{\sin^2(k r)}{(k r)^2}, & l\ \text{odd (even)} \\
    \frac{\cos^2(k r)}{(k r)^2}, & l\ \text{even (odd)}
    \end{cases}
\end{eqnarray}

The dipole components of $\vc{N}_{klm}$ for $l=1$, $m=0$ have the limiting behaviour in the near field given by
\begin{eqnarray}
\label{eqn:Nnearfield}
\sqrt{6 \pi}\  \vc{N}_{k10}(\vc{r}\rightarrow 0) & = & \hat{z} \\
    \sqrt{6 \pi}\ ^+\vc{N}_{k10}(\vc{r}\rightarrow 0) & = & \frac{3}{2 } \frac{1}{i ( k r)^3}\left( 3\hat{r}(\hat{z}\cdot\hat{r})- \hat{z} \right),\label{eqn:Nplusnearfield}
\end{eqnarray}
and in the far field 
\begin{eqnarray}
\label{eqn:Nfarfield}
 \sqrt{6 \pi}\  \vc{N}_{k10}(\vc{r}\rightarrow \infty) & = & \frac{3}{2} \frac{\sin{k r}}{k r}\left( \hat{r}\times \hat{z} \times \hat{r}\right)\\
  \label{eqn:Nplusfarfield}
    \sqrt{6 \pi}\ ^+\vc{N}_{k10}(\vc{r}\rightarrow \infty) & = &  \frac{3}{2} \frac{ e^{i k r} }{i k r}\left( \hat{r}\times \hat{z} \times \hat{r}\right)
\end{eqnarray}
(To simplify the discussion we take $\vc{E}_d$ and $\vc{E}_i$ to be linearly polarized along $\hat{z}$. However, arbitrary polarization can be treated in a similar way by including the $m=\pm1$ components of $\vc{N}_{klm}$.) The source-free solutions for all other $\vc{N}_{klm}(\vc{r})$ with  $l\neq 1$ vanish in the limit $r\rightarrow 0$, as well as all $\vc{M}_{klm}(\vc{r})$ for any $l$. This fact has an important consequence: the dipole component for an arbitrary $\vc{E}_i(\vc{r})$, $\vc{B}_i(\vc{r})$ is entirely and solely determined by the value of $\vc{E}_i(0)$ at the origin.

Consider a dipole at the origin subject to an excitation field $\vc{E}_i(\vc{r},t)=E_0 \vc{g}(\vc{r}) e^{- i \omega t}$, where the field at $r\rightarrow0$ is characterized by $E_0$ and $\vc{g}(0)=\hat{z}$: $\vc{E}_i(0,t)=E_0 \hat{z} e^{-i \omega t}$. We assume that $E_0$ satisfies the slowly varying approximation $|\dot{E}_0(t^\prime)| \ll \omega |E_0(t^\prime)|$, but $\vc{E}_i(\vc{r},t)$ is otherwise an arbitrary solution to Maxwell's equations. Comparing to Eq.~\ref{eqn:Eexpansion} and~\ref{eqn:Nnearfield}, $\vc{g}(\vc{r}) = \vc{g}_{\mathrm{d}}(\vc{r}) + \vc{g}_{\perp}(\vc{r})$, where $\vc{g}_{\mathrm{d}}(\vc{r})=\sqrt{6 \pi} \vc{N}_{k10}(\vc{r})$ and $\vc{g}_{\perp}(\vc{r})$ is orthogonal to $\vc{N}_{k10}(\vc{r})$.

We define a characteristic field strength associated with the dipole $d$ to be $E_d= \omega^3 d/6\pi \epsilon_0 c^3$, where for the classical calculation $d=\frac{1}{2}\alpha E_0$ and in the quantum case $d=d_{ij}\hat{\sigma}_-(t^\prime)=|\bra{j} e \vc{r} \ket{i}|\hat{\sigma}_-(t^\prime)$. 
The dipole radiated field can be written $\vc{E}_d(\vc{r},t) = E_d \vc{f}(\vc{r})e^{-i \omega t}$ where the far-field bahavior of $\vc{f}$ is 
\begin{equation}
    \vc{f}(\vc{r}) \simeq \frac{3}{2} \frac{e^{i k r}}{k r} \left(\hat{r} \times  \hat{z} \times \hat{r}\right) .
    \label{eqn:fdef}
\end{equation}
Comparing eq. (\ref{eqn:Nplusfarfield}) and Eq. (\ref{eqn:fdef}) we find 
$\vc{f}(\vc{r})=i \sqrt{6 \pi}\ ^+\vc{N}_{k10}(\vc{r})$. The dipole structure of $\vc{f}(\vc{r},t)$ and
$\vc{g}_{\mathrm{d}}(\vc{r},t)$ determines the spatial dependence of the radiated energy.

Poynting's theorem describes the flow of energy within a closed surface $\Omega$ containing the molecule. Within the quasistatic, slowly varying approximation, it relates $P_{\mathrm{mol}}$, the rate of change of the internal energy of the molecule, to  $\Delta \vc{S}= \vc{S}_{\mathrm{tot}}-\vc{S}_{i}$, the difference between the Poynting vector with and without the molecule: 
\begin{equation}
    P_{\mathrm{mol}} = -  \int_{\Omega} \Delta \vc{S} \cdot d\vc{\Omega}.
\label{eqn:Poynting}
\end{equation}
where $\vc{S}= (1/\mu_0) \vc{E}\times \vc{B}$. Choosing the surface to be a sphere of radius $r$, $d\vc{\Omega}=\hat{\vc{r}} d \Omega$, the orthogonality Eq.~\ref{eqn:crossprodorth} means that $\vc{g}_{\perp}(\vc{r})$ drops out of the integral in Eq.~\ref{eqn:Poynting}, and $P_{\mathrm{mol}}$ is independent of $\vc{E}_i(\vc{r},t)$ other than its value at $\vc{r} = 0$.
Note that the properties of $\vc{N}_{klm}$ and $\vc{M}_{klm}$ listed above (eqns.~\ref{eq:sourcednorm} and \ref{eq:mixednorm}) also imply that the time-averaged far-field solid angle integral of $c$ times the energy density difference, $c \Delta u = c \epsilon_0( \vc{E}_i \cdot \vc{E}^*_d+ \frac{1}{2}|\vc{E}_d|^2) $ is equal to the solid angle integral of $\Delta \vc{S}\cdot \hat{\vc{r}}$.
Therefore the cycle averaged integral $P_{\mathrm{abs}}=-P_{\mathrm{mol}}$ can be written in terms of (the real part of) $c \Delta u$
\begin{eqnarray}
  P_{\mathrm{abs}}&=& c \int r^2\Delta u d\Omega \nonumber \\
  &=& c \epsilon_0  \int r^2 E_0 E_d \vc{g}^*_{\mathrm{d}}\cdot \vc{f} +\frac{1}{2}|E_d|^2 | \vc{f}|^2 \ d \Omega.
 \end{eqnarray}
Using the expressions for $\vc{g}_d(\vc{r})$, $\vc{f}(\vc{r})$ and $E_d$, and performing the integrals gives
\begin{equation}
    P_{\mathrm{abs}}= \omega\ \mathrm{Im}[d E_0]+ \omega\ d E_d . \\
    \label{eqn:Pextexplicit}
\end{equation}
Note that with $\vc{d}(t)=\vc{d} e^{-i \omega t}$ the first term in Eq.~\ref{eqn:Pextexplicit} is equal to $P_{\text{ext}}= -P_d = - \avg{\vc{E}_i(0)\cdot \dot{\vc{d}}}{t}$. 



\subsection{Optical Theorem in the vector spherical function basis}
The optical theorem relates the imaginary part of the forward scattering amplitude to the total cross section of the scatterer, and is equivalent to eq. (\ref{eqn:sigmaAbs}) in the linear polarizable regime. The relation is required by energy conservation, and for a dipole scatterer, it relates the integral of the interference term eq. (\ref{eq:Scross}) to the energy dissipated by the dipole.
It is usually derived using incident plane wave excitation within the paraxial limit by integrating $\vc{S}_{\mathrm{cross}}$ over the surface of a far screen \cite{jackson_classical_1999}. The relationship between the local dipole response and the far field scattering distribution should be independent of the spatial form of $\vc{E}_i$.

For example, if one assumes an incident gaussian beam focused on the position of the dipole, where both $\vc{E}_i$ and $\vc{E}_d$ are radial waves in the far field $\propto e^{i k r}/r$, integration of eq. \ref{eq:Scross} over a spherical surface should equal the energy dissipated. It turns out that the Gouy phase is required in order for $\vc{E}_i(\vc{r}=0)$ to have the correct phase relation with $\vc{d}$ to conserve energy. In general, the phase of $\vc{E}_i(\vc{r}=0)$ must be related to the far-field radiation pattern.

\begin{figure}[H]
    \centering
    \includegraphics[scale=1]{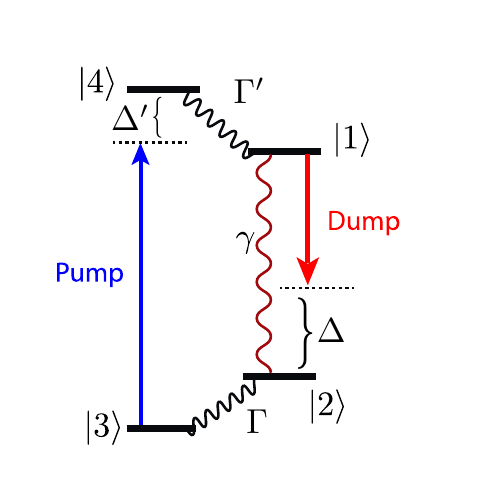}
    \caption{Four-level system energy diagram as simple model of a pumped fluorophore.}
    \label{fig:4leveldiagram}
\end{figure}

\subsection*{Four level quantum model}
To understand qualitatively the impact of spontaneous emission on fluorophore scattering, we simplify the description of our fluorophore to a four level model, as shown in fig.~\ref{fig:4leveldiagram}. Although simple, this model captures many of the features discussed in the main text. A pump beam with Rabi frequency $\hbar \Omega_p=\vc{d}\cdot \vc{E}_p$ excites the dipole to an excited state $|4\rangle$, which decays rapidly (at rate $\Gamma^\prime$) to a different excited  state $|1\rangle$ with much longer lifetime $\tau = \gamma^{-1}$, where $\gamma = \omega^3 d_{12}^2/ 3 \pi \epsilon_0 \hbar c^3$ is the radiative decay rate of $|1\rangle$.  A dump beam, with Rabi frequency $\hbar \Omega_i=\vc{d}\cdot \vc{E}_i$ and detuning $\Delta$, drives the dipole to a lower state $|2\rangle$, which quickly decays non-radiatively back to the ground state $|3\rangle$ at rate $\Gamma$. The radiation of interest arises from the $|1\rangle \rightarrow |2\rangle$ transition. The non radiative relaxation rates are taken to be much larger than the radiative rate, $\Gamma, \Gamma^\prime \simeq 1000 \gamma$, typical of fluorophores.

Solving for the radiated field of a single driven quantum dipole in the semi-classical approximation involves first solving the master equation for the dynamics of the dipole, given the incident field. (We work in the Heisenberg  picture, where the operators are time dependent.) We note that since the system is strongly dissipative, the populations can be described by incoherent, classical rate equations~\cite{hell1994breaking}. The Rayleigh scattering, however, is a coherent process and  depends on the coherence between states $|1\rangle$ and $|2\rangle$.

Given the solution for $\hat{\vc{d}}(t)$, the radiated field from a transition between states $|j\rangle \rightarrow |i\rangle$ is given by Eq.~\ref{eqn:quantdip}.
Expectation values of operators $\langle \hat{O}(t)\rangle = {\text{tr}}\left[ \hat{\rho}\hat{O}(t)\right]={\text{tr}}\left[ \hat{\rho}(t)\hat{O}(0)\right]$ are determined using solutions $\hat{\rho}(t)$ to the master equation
\begin{equation}
    \partial_t \hat{\rho}=\frac{- i}{\hbar}\left[H,\hat{\rho}\right]- \sum_i\frac{\gamma_i}{2}\left(C_i^\dagger C_i \hat{\rho} +\hat{\rho} C_i^\dagger C_i- 2 C_i \hat{\rho} C_i^\dagger\right).
\end{equation}
Here, $H$ is the semiclassical Hamiltonian governing the system and $C_i$ are the dissipative jump operators with associated decay rates $\gamma_i$. 

For the four-level model considered here,
\begin{align}
    H = &\Delta |2\rangle \langle 2|+ \left( \frac{\Omega_i}{2} |1\rangle \langle 2|+h.c. \right)\nonumber \\ 
     & + \Delta^\prime |4\rangle \langle 4|+ \left( \frac{\Omega_p}{2} |3\rangle \langle 4| + h.c. \right),
\end{align}
where $\Delta$ and $\Delta'$ are the detunings from the excited and vibrationally excited ground state respectively, $\Omega$ are the Rabi frequencies and the jump operators/decay rates are given by
\begin{align}
    C_1=|1\rangle \langle 4|&\ \ \  \gamma_1 =\Gamma^\prime, \\
    C_2=|3\rangle \langle 2|&\ \ \  \gamma_2 =\Gamma, \\
    C_3=|2\rangle \langle 1|&\ \ \  \gamma_3 =\gamma. 
\end{align}
We will consider resonant pumping, $\Delta^\prime =0$ for the remainder of this section.

The far-field radiated power 
due to the stimulated emission and isotropic dipole terms, the quantum version of Eq.~\ref{eqn:Pextexplicit}, are given by
\begin{eqnarray}
P_{\mathrm{stim}} &=& \hbar \omega \ \mathrm{Im}\left[ \Omega^*_{i} \ \rho_{12} \right] \\
P_{4 \pi}& = & \hbar \omega \ \gamma \  \rho_{11}
\end{eqnarray}
where 
$\rho_{i j}$ are the elements of the density matrix for the four level system, determined by solving the master equation governing the system. In the limit $\Gamma, \Gamma^\prime  \gg \Omega_p, \Omega_i, \gamma$, the steady state solutions for $\rho_{i j}$ give
\begin{eqnarray}
P_{\mathrm{stim}} &=&\hbar \omega \ \Gamma_{{d}}\ \frac{\Gamma_{{p}}}{\Gamma_{{p}}+\Gamma_{{d}}+\gamma}\\
P_{4 \pi}& = &\hbar \omega \  \gamma\ \frac{\Gamma_{{p}}}{\Gamma_{{p}}+\Gamma_{{d}}+\gamma},
\end{eqnarray}
where the effective pump and dump rates are 
\begin{eqnarray}
\Gamma_{{p}}&=& \left( \frac{\Omega_p^2}{\Gamma^\prime}\right)
\nonumber \\
\Gamma_{{d}}&=&  \left( \frac{\Omega_i^2}{\Gamma}\right) \frac{1}{1+4(\Delta/ \Gamma)^2 }
\end{eqnarray}
The ratio of total 4$\pi$ scattering to stimulated emission is $P_{4\pi}/P_{\mathrm{stim}} =\gamma /\Gamma_d$, which vanishes in the limit of strong dumping, $\Gamma_d \rightarrow \infty$. 

The part of the total $P_{4 \pi}$ that is coherent with the dump field~\cite{Scully1997} (i.e. the excited state Rayleigh scattering described in the classical discussion above) is
\begin{eqnarray}
P_{{R}}&=&  \hbar \omega\  \gamma\  \left|  \rho_{12} \right| ^2 \nonumber \\
&=& \hbar \omega \  \gamma\ \frac{\Gamma_{{d}}}{\Gamma} \left( \frac{\Gamma_{{p}}}{\Gamma_{{p}}+\Gamma_{{d}}+\gamma}\right) ^2. 
\label{eqn:PRay}
\end{eqnarray} 
For typical strong driving parameters $ \Gamma \gg \Gamma_d \gg \gamma$, the fraction of the total scattering that is coherent Rayleigh scattering is small, $P_R/P_{4 \pi} \leq \Gamma_d / \Gamma $, and is yet smaller than the power in the stimulated interference term, $P_R/P_{\mathrm{stim}}\leq \gamma/\Gamma$. The only way to increase $P_R$ to near the maximum allowed $\hbar \omega \gamma$ is to dump at a rate approaching the fast relaxation time scale $\Gamma$

In the strongly pumped, fully inverted limit, $\Gamma_{{p}}\gg \Gamma_{{d}}$ and $\rho_{11} \simeq 1$, $P_{4 \pi}$ is fixed to its maximum value, $P_{4 \pi}=\hbar \omega \gamma$, independent of the dump rate $\Gamma_{{d}}$. Increasing the dump field does not increase the dipole radiated power $P_{4 \pi}$. In fact, outside of the strongly pumped limit, the only effect that the dump field has on $P_{4 \pi}$ is through the lowering of the equilibrium excited state population $\rho_{11}$, and $P_{4 \pi}$ actually {\em decreases} with increasing $\Gamma_{{d}}$. 

In the weakly dumped, linear excitation regime, the  coherent part of the $|1\rangle \rightarrow |2\rangle$ dipole response to the incoming field $\vc{E}_i$ can be described by a polarizability, $\langle \vc{d}(t) \rangle= \vc{d}_{12} \rho_{12} =  \alpha \vc{E}_i/2$. Writing the electric field magnitude $E_i$ in terms of $d_{12}$, $\Omega_i$ and $\gamma$ as $d_{12} \Omega_i/ \gamma=(3 \epsilon_0 \lambda^3/8 \pi^2) E_i$ gives the explicit expression for $\alpha_{\mathrm{inv}}$ in the fully inverted limit:
\begin{eqnarray}
      \alpha_{\mathrm{inv}} &=& \frac{3 \epsilon_0 \lambda^3}{4 \pi^2}\gamma \ \frac{-i}{ \Gamma+2 i \Delta} \nonumber \\ 
      &=& \frac{3 \epsilon_0 \lambda^3}{4 \pi^2}\ \frac{\gamma}{\Gamma} \   \frac{-2\Delta/\Gamma-i }{1+4\Delta^2/\Gamma^2} \
\end{eqnarray}
Using eqn.~\ref{eqn:sigmaAbs} and eqn.~\ref{eqn:sigmaR}, we confirm that $\sigma_R/\sigma_{\mathrm{stim}}=\gamma/\Gamma \ll 1$.

As a side note, comparing to the polarizability of a non-inverted 2-level system with the same dipole matrix element $d_{12}$ and a non-radiative relaxation rate $\Gamma\gg \gamma$,
\begin{equation}
    \alpha_{\mathrm{2lev}}=\frac{3 \epsilon_0 \lambda^3}{4 \pi^2}\   \frac{\gamma}{\Gamma}\  
    \frac{2\Delta/\Gamma+i }{1+4\Delta^2/\Gamma^2},
\end{equation}
the inverted polarizability has opposite sign and equal magnitude.



This 4-level model ignores contributions to the polarizability from additional excited levels, which can in principle provide larger scattering than non-inverted, ground state systems~\cite{Crowell2018}.

\bibliography{StimRefs.bib}

\end{document}